\documentclass[aps,reprint,superscriptaddress, longbibliography]{revtex4-1}

\usepackage{amsmath}
\usepackage{amstext}
\usepackage{amssymb}
\usepackage{graphicx}
\usepackage{import}
\usepackage{appendix}
\usepackage[pdfpagemode=UseNone,pdfstartview=FitH,colorlinks=true,linkcolor=blue,citecolor=blue,urlcolor=blue]{hyperref}
\usepackage[all]{hypcap}
\usepackage{xcolor}

\begin{document}

\title{Fast, Lifetime-Preserving Readout for High-Coherence Quantum Annealers}

\newcommand{\ngc}{Northrop Grumman Corporation, Linthicum, Maryland 21090, USA}
\newcommand{\usc}{University of Southern California, Los Angeles, California 90089, USA}

\author{Jeffrey A. Grover}
\author{James I. Basham}
\author{Alexander Marakov}
\author{Steven M. Disseler}
\author{Robert T. Hinkey}
\author{Moe Khalil}
\author{Zachary A. Stegen}
\author{Thomas Chamberlin}
\author{Wade DeGottardi}
\author{David J. Clarke}
\author{James R. Medford}
\author{Joel D. Strand}
\author{Micah J. A. Stoutimore}
\author{Sergey Novikov}
\author{David G. Ferguson}
\affiliation{\ngc}
\author{Daniel Lidar}
\affiliation{\usc}
\author{Kenneth M. Zick}
\author{Anthony J. Przybysz}
\affiliation{\ngc}

\begin{abstract}
We demonstrate, for the first time, that a quantum flux parametron (QFP) is capable of acting as both isolator and amplifier in the readout circuit of a capacitively shunted flux qubit (CSFQ).
By treating the QFP like a tunable coupler and biasing it such that the coupling is off, we show that $T_1$ of the CSFQ is not impacted by Purcell loss from its low-Q readout resonator ($Q_e = 760$) despite being detuned by only $40$ MHz.
When annealed, the QFP amplifies the qubit's persistent current signal such that it generates a flux qubit-state-dependent frequency shift of $85$ MHz in the readout resonator, which is over $9$ times its linewidth.
The device is shown to read out a flux qubit in the persistent current basis with fidelities surpassing $98.6\%$ with only $80$ ns integration, and reaches fidelities of $99.6\%$ when integrated for $1$ $\mu$s.
This combination of speed and isolation is critical to the readout of high-coherence quantum annealers.
\end{abstract}

\maketitle

\section{INTRODUCTION}
\label{sec:intro}
Quantum annealing is a heuristic algorithm typically employed for solving optimization problems formulated in terms of finding ground states of classical Ising spin Hamiltonians~\cite{Kadowaki1998}, closely related to adiabatic quantum computing~\cite{Farhi2001}.
In recent years it has generated a great deal of activity and interest, both theoretical and experimental~\cite{Albash2018, Hauke2020}.
Improvements in device technology are essential in order to approach the point where such devices can surpass the power of classical computing.
Crucially, in order to achieve advantage on computational problems of interest, quantum annealers must harness coherent quantum effects~\cite{Childs2001, Albash2015}.

Architectures for quantum annealers commonly employ superconducting circuits~\cite{Johnson2011, Leib2016, Puri2017, Onodera2020}, and a particularly promising route uses tunable flux qubits~\cite{Paauw2009, Harris2010, Zhu2010, Fedorov2010, Gustavsson2011} that have small persistent currents~\cite{Novikov2018} in order to take advantage of quantum coherence.
It has been shown, both in flux qubits~\cite{Yan2016, Weber2017, Novikov2018} and fluxonium~\cite{Manucharyan2009, Nguyen2019}, that decreasing the magnitude of the qubit persistent current, $I_p$, is a key factor in achieving long coherence times.
This constraint is in conflict with the need for having fast, high-fidelity readout of the flux qubit states, which are encoded in the sign of $I_p$ at the end of the annealing protocol.
Generating the necessary coupling to a resonator or superconducting quantum interference device (SQUID) for readout, given the small $I_p$, requires the construction of large mutual inductance transformers, which can be prohibitive in a high-coherence fabrication process. 
Furthermore, if directly connected for readout, then overcoupling to the environment limits the qubit's coherence.
Current relatively low-coherence commercial quantum annealers utilize a quantum flux parametron (QFP) to lock-in the qubit's state at readout~\cite{Harris2010} and it has been shown that this type of readout can be scaled to thousands of qubits~\cite{Whittaker2016}.
However, it remains an open problem to show that this type of readout does not limit qubit coherence.
We assert that not only can a QFP amplify and lock the qubit circulating current state, but also that its biases can be adjusted such that it isolates the qubit from the rest of the readout chain during quantum annealing.
Although the concept dates back to some of the earliest proposed flux qubit designs~\cite{Clarke2002}, it so far has neither been demonstrated experimentally, nor has it been adapted for use with resonators rather than dc SQUIDs. 
This work demonstrates fast, high-fidelity readout using a QFP and tunable resonator, and shows how the QFP acts as a tunable coupler to protect the flux qubit from Purcell loss through the tunable resonator.

Our demonstration vehicle is a high-coherence flux qubit connected via an intermediary QFP to a tunable resonator. 
The tunable resonator is created by incorporating an rf SQUID into the current antinode of a quarter-wave resonator such that the frequency is sensitive to dc flux and is therefore able to sense the circulating current state of the QFP. 
Counting on the improved isolation provided by the QFP, the readout resonator is designed with low external quality factor, $Q_e$, to facilitate fast interrogation of the resonant frequency.
The QFP is shown to substantially amplify the flux shift from the flux qubit into the tunable resonator. 
The full range of this flux shift, detunes the resonator by many  linewidths and enables single-shot readout with good fidelities even for integration times less than $100$ ns.
The isolation provided by the QFP is demonstrated both spectroscopically and with time-domain measurements of flux qubit lifetime, $T_1$.
With the QFP biased in isolation mode the qubit frequency can be tuned to within a few tens of megahertz of the tunable resonator without suffering measurable degradation in the $T_1$ time.
The QFP provides the isolation and amplification necessary for fast, high-fidelity readout without impacting qubit lifetime.

\section{DEVICE DESIGN AND THEORY OF OPERATION}
\label{sec:theory}
\subsection{Device overview}
\label{subsec:device}
\begin{figure*}[t]
	\begin{center}
		\includegraphics[width=\textwidth]{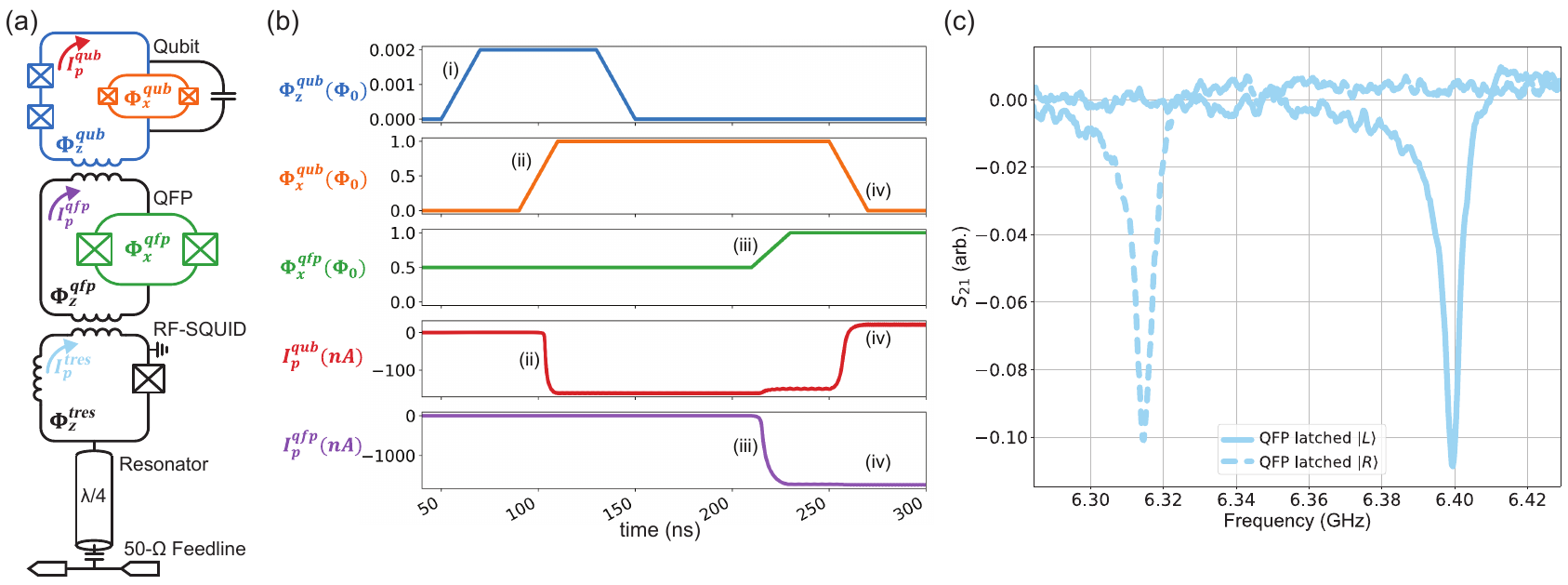}
    \end{center}
	\caption{
    (a) Schematic of device (for more details see Appendix~\ref{sec:device}).
    A four-junction CSFQ is inductively coupled to a QFP via their $z-$flux loops.
    The QFP is further coupled to a rf SQUID loop attached to the current antinode of a $\lambda/4$ resonator.
    Relevant fluxes and persistent currents are labeled and color coded.
    Each loop flux is controllable via an independent external bias line (not drawn).
    Dispersive diagnostic resonators coupled to the CSFQ and QFP are omitted from the schematic.
    (b) Example control sequence (top three plots) and the resulting persistent currents modeled in WRspice (bottom two plots).
    Applying a small amount of $\Phi^{qub}_z$ causes an initial tilt to the qubit potential (i).
    We anneal the qubit by increasing $\Phi^{qub}_x$ to a value of 1~$\Phi_0$, inducing a nonzero persistent current in the qubit (ii).
    After the qubit anneal is complete, we subsequently anneal the QFP, which senses $I^{qub}_z$ and latches to a much larger persistent current $I^{QFP}_z$ (iii).
    We can then reset the qubit by lowering its barrier, and the QFP remains latched (iv).
    Note the QFP persistent current causes some back action on the qubit, which causes it to return to a steady-state current greater than $0$ while the QFP remains annealed.
    (c) Representative plot of experimental tunable resonator line shifts, $S_{21}$, versus frequency.
    The tunable resonator position moves by multiple linewidths when the QFP latches to either a positive (light blue solid) or negative (light blue dashed) persistent current state.
	}
	\label{fig:schematic}
\end{figure*}

In Fig.~\ref{fig:schematic}(a) we depict a schematic of the device.
A capacatively shunted flux qubit (CSFQ)~\cite{You2007, Yan2016} with independent $z-$ and $x-$flux tunability~\cite{Novikov2018} is inductively coupled to the QFP, which is in turn inductively coupled to a tunable resonator.
The device was fabricated at MIT Lincoln Laboratory by patterning high-quality aluminum on a high-resistivity silicon substrate~\cite{Oliver2013}, and the qubit Josephson junctions are designed such that the qubit persistent current $I_p^{qub} \approx 170$ nA at the end of annealing to maintain high coherence.
The flux qubit $z$ loop is coupled to the QFP $z$ loop with a geometric mutual inductance, and the same is true of the QFP $z$ loop to tunable resonator rf SQUID loop coupling.
Each loop flux is controllable via an independent external bias line, supplied by a 1 GS/s arbitrary waveform generator (see Appendix~\ref{sec:fridge} for more wiring information).
We refer the reader to Table~\ref{tab:device} in Appendix~\ref{sec:device} for a detailed list of device parameters and to Appendix~\ref{sec:tres} for details about the tunable resonator.

In many ways, the QFP is a flux qubit with larger junction critical current.
It, too, has  $z-$ and $x-$flux controls that apply $\Phi^{QFP}_z$ and $\Phi^{QFP}_x$ (see Fig.~\ref{fig:schematic}(a)), respectively.
The larger critical currents and flux control allow the QFP to act as both tunable coupler and amplifier in this circuit.
The effective $\beta^{QFP}_L$ of the QFP, which sets the height of the barrier in the double-well potential, is tunable via its $x$ flux:
\begin{equation}
    \beta^{QFP}_L = \frac{4 \pi I^{QFP}_{c} L^{QFP}}{\Phi_0} \cos \left( \frac{\pi \Phi^{QFP}_{x}}{\Phi_0} \right)\,.
    \label{eq:beta}
\end{equation}
Here $I^{QFP}_{c}$ is the critical current of each junction in the $x$ loop, and $L^{QFP}$ is the linear loop inductance.
These two design parameters are chosen such that the maximum $\beta^{QFP}_L$ is 2.5, ensuring that the QFP's potential energy can be made double welled ($\beta^{QFP}_L > 1$).
By ramping $\Phi_x^{QFP}$ from 0 to 1~$\Phi_0$, the potential of the QFP can be transformed from single well ($0<\Phi_x^{QFP}<0.56$ $\Phi_0$) to double well ($0.56\ge\Phi_x^{QFP}<1$ $\Phi_0$), just as in a flux qubit.

Tuning $\beta^{QFP}_L$ changes the QFP's susceptibility, $\chi = dI_p/d\Phi_z^{QFP}$, and thus the effective mutual inductance, $M_{eff}$, between the flux qubit and tunable resonator according to~\cite{Harris2010}
\begin{equation}
    \chi = \frac{1}{L^{QFP}}\frac{\beta^{QFP}_L}{1+\beta^{QFP}_L}\,,
    \label{eq:chi}
\end{equation}
\begin{equation}
    M^{eff} = M^{qub}M^{tres}\chi\,,
    \label{eq:Meff}
\end{equation}
where $M^{qub}$ and $M^{tres}$ are the qubit-to-QFP and QFP-to-tunable resonator mutual inductances, respectively.
Therefore, the QFP can be used to isolate the flux qubit from spontaneous emission due to the Purcell effect during the annealing operation in the following manner: setting $\Phi_x^{QFP} = \Phi_0/2$, so that $\beta^{QFP}_L = 0$ by Eq.~(\ref{eq:beta}), causes the susceptibility of the QFP to vanish by Eq.~(\ref{eq:chi}).

The QFP acts as a current amplifier by matching the qubit's direction of circulating current, but with a much larger magnitude.
This amplification depends on the strength of $M^{qub}$ and the width of the QFP transition, as described below.
The QFP's probability of ending up with a positive or negative circulating current state is well approximated by
\begin{equation}
    P_{R(L)} = \frac{1}{2} \left[ 1 - \tanh \left( \frac{\Phi^{QFP}_z - \Phi^{qub}_{R(L)}}{w} \right) \right]\,,
    \label{eq:tanh}
\end{equation}
where the width of the transition, $w$, is limited by noise in $\Phi_z^{QFP}$ and nonadiabatic transitions that occur during the QFP annealing process.
The term $\Phi^{qub}_{R(L)}$ represents the flux coupled into the QFP $z$ loop from the flux qubit when it is in its right(left) circulating current state.
When the flux qubit is not in a persistent current state, that term is considered to be 0. 
The graphical representation of the QFP transition is the plot of Eq.~(\ref{eq:tanh}) as $\Phi^{QFP}_z$ is raised from an initial value much smaller than $0$ to a final value much bigger than $0$, which forms an ``$s$ curve''~\cite{Khezri2020} when the initial and final values of $\Phi^{QFP}_z$ are sufficiently far from 0. 
Aggressive filtering of bias lines and slow annealing rates (approximately megahertz) have been shown to give QFP widths as low as $w\sim100$ $\mu\Phi_0$~\cite{Harris2009}.
By designing $\Delta\Phi = 2I_p^{qub}M^{qub}\gg w$, the signal from the flux qubit can be amplified by a factor of $I_p^{QFP}/I_p^{qub}$, where $I_p^{QFP}$ is the persistent current in the QFP when it is annealed (Fig.~\ref{fig:schematic}(b)). 
In this case the ratio $I_p^{QFP}/I_p^{qub}$ is simulated to be about $10$.

\subsection{Readout protocol}
\label{subsec:readout protocol}
The flux qubit is read out by manipulating the biases on the flux qubit, QFP (Fig.~\ref{fig:schematic}(b)), and tunable resonator so as to map the circulating current state of the flux qubit onto a change in transmission through the feedline driving the tunable resonator (Fig.~\ref{fig:schematic}(c)).
The persistent currents in the flux qubit and QFP, as simulated in WRspice~\cite{WRspice}, are depicted in the bottom two subplots of Fig.~\ref{fig:schematic}(b).
At $t=0$, the flux qubit control loop biases are set such that $\Phi_x^{qub} = 0$ and $\Phi_z^{qub} = 0$, accounting for any unintended flux offsets caused by trapped flux.
The QFP $x$ loop is biased such that $\Phi_x^{QFP} = \Phi_0/2$ in order to suppress the persistent current in the QFP and isolate the flux qubit by minimizing its susceptibility to external flux, $\Phi_z^{QFP}$.
The QFP $z$ loop is biased to  $\Phi_z^{QFP} = 0$, again, accounting for any unintended flux offsets, such that it will be sensitive to the flux qubit circulating current state.
The first step is to anneal the flux qubit by raising $\Phi_x^{qub}$ from 0 to $\Phi_0$, which causes it to latch into either a left or right circulating current state, as shown in Figs.~\ref{fig:schematic}(b)(i) and (ii).
Next, the QFP is annealed in a similar fashion such that it latches into a circulating current state that depends on the state of the flux qubit, as in Fig~\ref{fig:schematic}(b)(iii).
The QFP is designed to apply a state-dependent flux to the tunable resonator of  $\pm50$ m$\Phi_0$.
The tunable resonator is biased to $\Phi_z^{tres} = \Phi_0/4$ in order to increase its sensitivity to flux such that the $\pm50$ m$\Phi_0$ shift results in an $85$ MHz frequency shift, which is $9$ times the linewidth.
We note that this large state-dependent shift is achieved while only utilizing $2\%$ of the qubit loop inductance for the readout mutual inductance.
Finally, the tunable resonator is interrogated by measuring the transmission through a feedline that is strongly coupled to it with $Q_e = 760$.
The probe frequency is set to be near one of the QFP-state-dependent resonances (see Fig.~\ref{fig:schematic}(c)) so that the signal amplitude contains the state information.
Once the readout signal is encoded in the QFP persistent current, it is protected from any errors due to flux qubit tunneling or thermal excitations, and can be integrated for as long as is necessary to facilitate single-shot readout. 

\section{READOUT PERFORMANCE}
\label{sec:readout performance}
\begin{figure*}[t]
    \begin{center}
        \includegraphics[trim={0 2cm 0 2cm}, width=\textwidth]{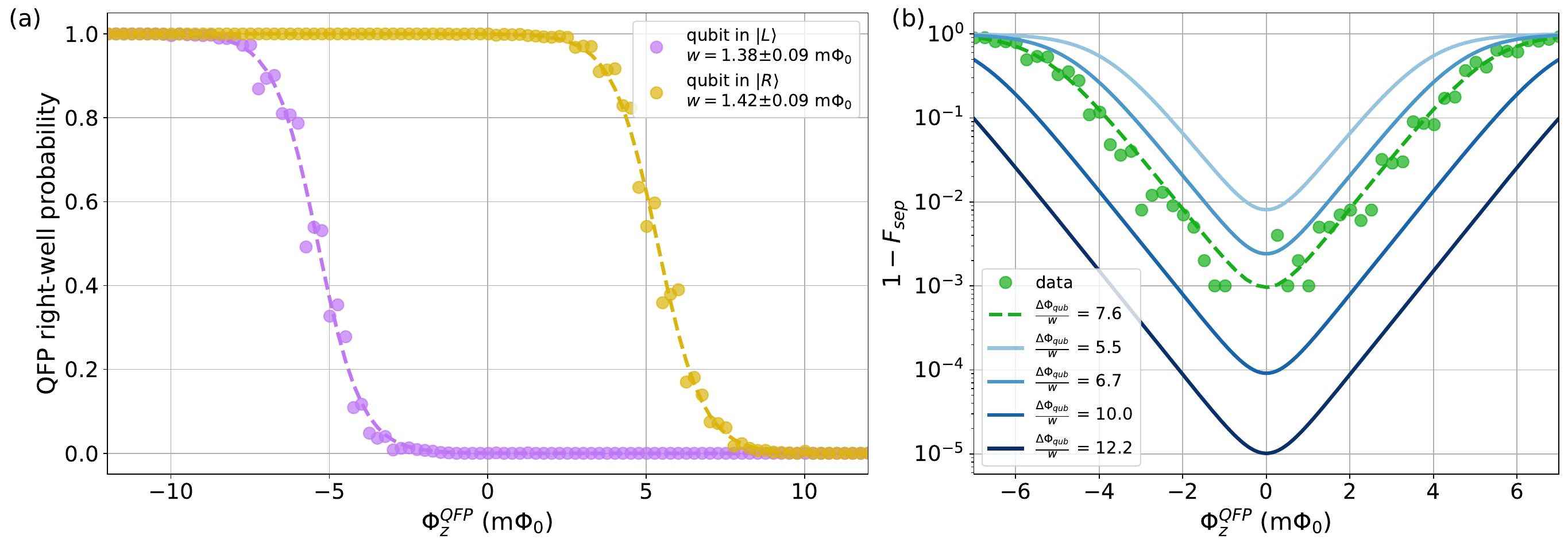}
    \end{center}
    \caption{(a) Measured (circles) and fitted (dashed lines) QFP $s$ curves for flux qubits prepared in the left (purple) and right (yellow) circulating current states.
     Widths extracted from Eq.~(\ref{eq:tanh}) are found to be $1.38$ and $1.42$ m$\Phi_0$ for the left and right flux qubit persistent current states, respectively.
     (b) QFP s-curve separation readout fidelity is plotted as $1 - F_{sep} = 1 - \left[P_R(\Phi_z) - P_L(\Phi_z)\right]$ for clarity.
     The fidelities are found by subtracting the QFP s-curve data (green circles), s-curve fits (green dashed), and a collection of theoretical curves with varying $\Delta\Phi^{qub}/w$ (solid lines).
     The s-curve separation limit on readout fidelity for the measured device is found to be $99.91\% \pm 0.095\%$.
    }
    \label{fig:s-curve-fidelity}
\end{figure*}

QFP s-curve separation fidelity expressed as the ratio between the qubit flux signal, $\Delta\Phi^{qub}$, and the fitted QFP width, $w$, provides a natural and intuitive framework for evaluating the readout quality of this class of device.
Devices with poor $s$-curve separation will have their readout fidelity limited by the $s$ curves of the QFP.
In Fig.~\ref{fig:s-curve-fidelity}(a) we illustrate this principle with QFP $s$-curve data collected when the flux qubit is in a left circulating state (purple trace) and again when in a right circulating state (yellow trace).
It is immediately clear that we will achieve good separation readout fidelity, as ($\Delta\Phi^{qub} = 10.72$ m$\Phi_0$) is large compared to the averaged QFP widths ($w \sim 1.40$ m$\Phi_0$).

We calculate separation readout fidelity, $F_{sep}(\Phi_z) = P_R(\Phi_z) - P_L(\Phi_z)$, to be the difference between the two $s$ curves.
Readout error is defined as $1 - F_{sep}$ and is plotted in Fig.~\ref{fig:s-curve-fidelity}(b) to identify its theoretical lower bound. 
From the subtraction of the s-curve fit lines, the maximum separation readout fidelity is $99.91 \pm 0.095\%$ and the qubit flux signal to QFP width ratio is $7.65$.

Commercial annealer technologies have demonstrated single-qubit readout error probabilities of $10^{-5}$ and stated a goal of hitting $10^{-6}$ for problems with thousands of variables~\cite{Whittaker2016}.
This error budget would ensure that processor-wide readout fidelity stays at the $99.9\%$ level for a 1000-qubit annealer.
The path to improving our readout fidelity to the state of the art, and beyond, is quite clear: either decrease the QFP s-curve width or increase the mutual inductive coupling between the flux qubit and the QFP.
We reach $99.999\%$ fidelity when $\Delta\Phi^{qub}/w = 12.2$ (dark blue trace), which corresponds to either a $37\%$ reduction in $w$ to $0.88$ m$\Phi_0$ from $1.40$ m$\Phi_0$, or an increase in qubit-to-QFP mutual inductance of $60\%$ to $104$ pH from $65$ pH.
These results provide an outline of the parameter space occupied by high-performance QFP readout of the flux qubit, which can be accessed by moderate increases to qubit-QFP coupling or moderate reductions in the QFP width. 
Indeed, attempting to address both the qubit signal and QFP width simultaneously would require only relatively minor changes in device design, while driving readout error down by nearly two orders of magnitude.

\begin{figure}[t]
	\begin{center}
		\includegraphics[width=\columnwidth]{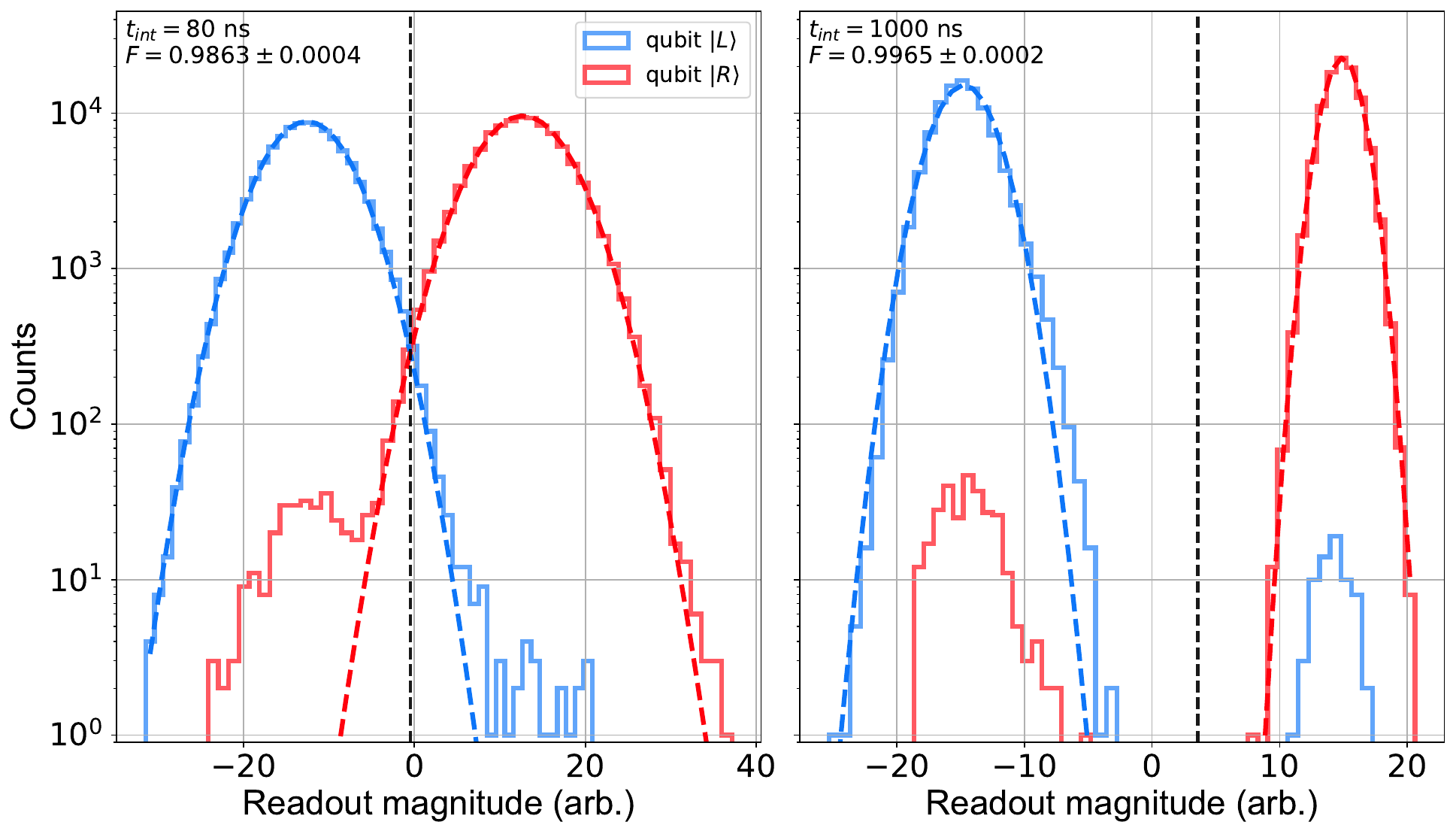}
	\end{center}
	\caption{
    Histograms of single-shot measurements of the qubit prepared in either its $|L\rangle$ (blue) or $|R\rangle$ (red) state for integration times of 80 ns (left) and 1 $\mu$s (right).
    Thick dashed lines are Gaussian fits to the dominant peaks, used to determine the overlap error.
    The vertical, black, dashed line indicates the threshold value, as determined by the point of intersection of the Gaussian fits.
    Fidelity improves from 98.63\% $\pm$ 0.04\% to 99.65\% $\pm$ 0.02\% as the integration time is increased from 80 ns to 1 $\mu$s.
    The overlap error correspondingly improves from 0.43\% to approximately $10^{-15}$\%, when the histograms become separated by more than 16$\sigma$.
	}
	\label{fig:histograms}
\end{figure}

To characterize the readout performance of the QFP, we perform measurement sequences similar to that outlined in Fig.~\ref{fig:schematic} and discussed previously.
Repeated single-shot measurements are collected for the qubit prepared in each opposing persistent current state, as well as for varied integration times.
Data for integration times of 80 ns and 1 $\mu$s are shown in Fig.~\ref{fig:histograms}, demonstrating clear distinguishability of the states.
A hard threshold is determined from the intersection point of the Gaussian fits to each peak (dashed lines in Fig.~\ref{fig:histograms}).
We define the overall measurement fidelity to be $F = 1 - \left[P(L \mid R) + P(R \mid L)\right]$~\cite{Gambetta2007, Walter2017}, where $P(x \mid y)$ is the probability of measuring the qubit in state $|x\rangle$ given it was prepared in state $|y\rangle$.
We achieve a fidelity of 98.63\% $\pm$ 0.04\% with only 80 ns of integration, improving to 99.65\% $\pm$ 0.02\% for the 1-$\mu$s case.
The overlap error from the Gaussian fits is found to be 0.43\% for 80 ns integration, and it becomes vanishingly small at longer integration times.

The gap between the observed fidelity of $99.65\%$ and the implied limit of $99.91\%$ has a likely experimental explanation.
We note that the misclassification errors visible in the right-hand plot of Fig.~\ref{fig:histograms} display an asymmetry where $P(L \mid R) > P(R \mid L)$.
If the $z$ flux for the QFP is not set to exactly zero, the overall separation fidelity is reduced (see Fig.~\ref{fig:s-curve-fidelity}(b)) and, morever, the errors are asymmetric depending on the direction of the offset (see Fig.~\ref{fig:s-curve-fidelity}(a)).
We estimate that an offset in $\Phi_z^{QFP}$ of approximately $1$ m$\Phi_0$ is enough to cause the observed misclassification asymmetry.
Residual uncalibrated linear crosstalk at about the $0.1\%$ level or nonlinear crosstalk due to junction asymmetry in the QFP $x$ loop~\cite{Khezri2020} are possible causes.
The design solutions outlined above for improving overall readout fidelity would also reduce the impact of such offsets.

We foresee clear paths to improving the readout performance by both raising fidelity limits and lowering integration time.
While fidelity limits can be improved by increasing CSFQ-to-QFP mutual inductance, the maximum speed of the readout is currently limited by the histogram overlap shown in Fig.~\ref{fig:histograms}. 
This overlap can be eliminated by increasing the microwave drive power of the tunable resonator, while reductions in $Q_e$ of the tunable resonator will also reduce the characteristic ringup time $1/\kappa$ of the resonator (currently $18$ ns), thereby lowering the minimum possible integration time. 
In fact, instrumentation improvements in the laboratory already afford an extra 10 dB of microwave drive, while $Q_e$ limits as low as 100 will be implemented in subsequent design iterations. 

\section{ISOLATION}
\label{sec:isolation}
The QFP isolates the flux qubit from environmental noise by turning off its coupling to the tunable resonator, and hence suppressing the Purcell effect.
This is demonstrated in two ways: the observed anticrossing of the flux qubit frequency with that of the tunable resonator and the effect of isolation on the flux qubit $T_1$ lifetime.

\begin{figure}[t]
	\begin{center}
		\includegraphics[width=\columnwidth]{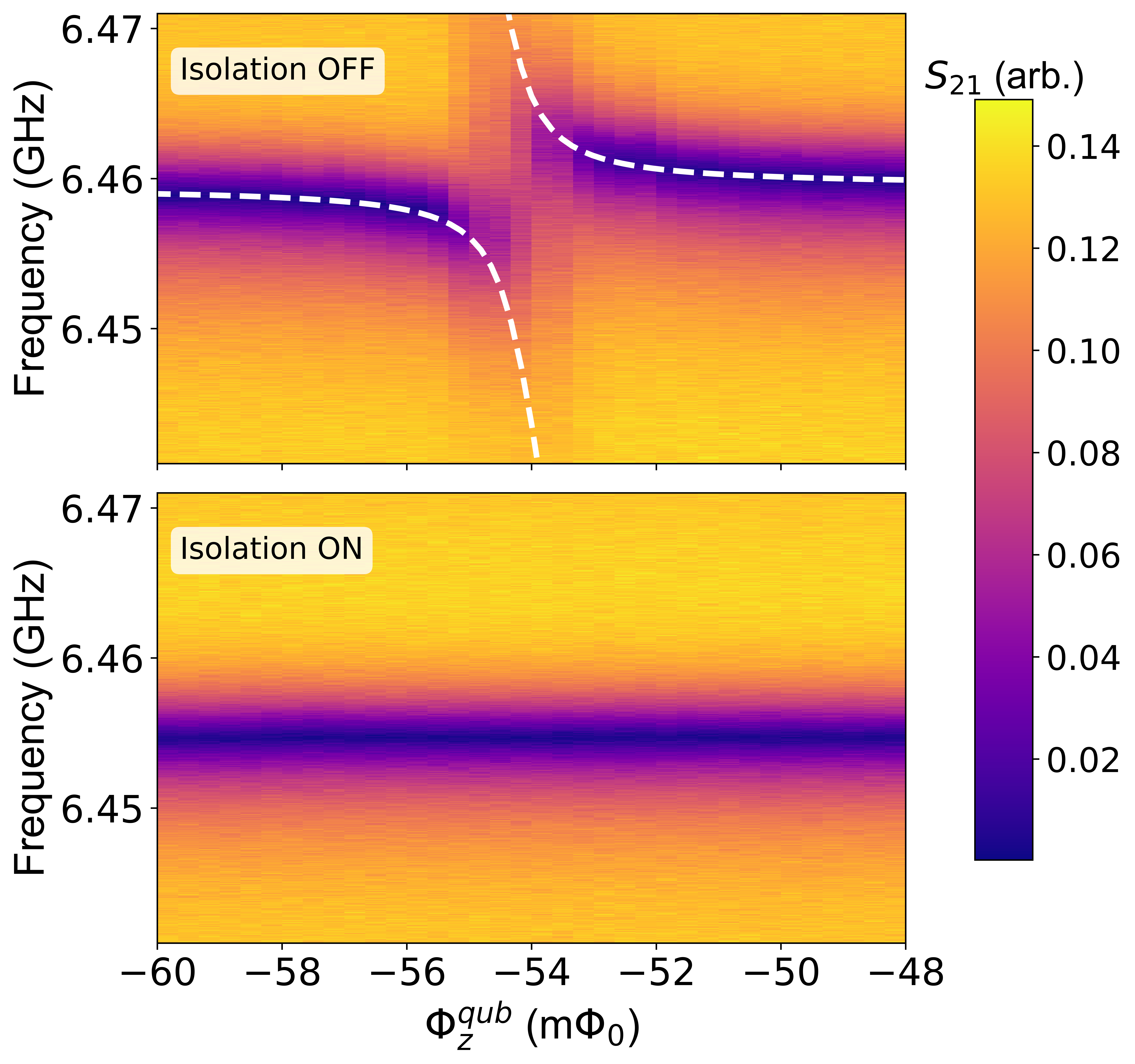}
	\end{center}
	\caption{
	Transmission ($S_{21}$) near the maximal tunable resonator frequency versus $\Phi^{qub}_{z}$, which tunes the qubit frequency through the resonator.
	The top plot has $\Phi^{QFP}_{x} = 0\,\Phi_0$, turning off the QFP's isolation.
	This produces an anticrossing from the qubit-resonator interaction.
	The dashed white lines are calculated from a full circuit Hamiltonian simulation (see Appendix~\ref{sec:circuitizer}) and correspond to an effective coupling of $g/2\pi = 9.8$ MHz.
	The bottom plot has $\Phi^{QFP}_{x} = 0.5\,\Phi_0$, turning on the isolation provided by the QFP.
	The anticrossing disappears because the qubit and resonator are no longer coupled.
	}
	\label{fig:isolation_anti}
\end{figure}

Recall from Eq.~(\ref{eq:Meff}) that the QFP can be seen as a tunable effective mutual inductance, $M_{eff}$, that couples the flux qubit $z$ loop to the tunable resonator rf SQUID loop.
In Fig.~\ref{fig:isolation_anti} we show how this effective mutual permits the flux qubit to interact with the tunable resonator.
When the QFP $x$ loop is biased to $0$~$\Phi_0$ and the tunable resonator is kept at its upper sweet spot frequency of $6.46$ GHz, a clear ant-crossing develops as the flux qubit frequency is swept through the resonator frequency (top plot of Fig.~\ref{fig:isolation_anti}).
A full circuit Hamiltonian simulation (see Appendix~\ref{sec:circuitizer}) verifies the existence and size of the anticrossing (dashed white lines in the top plot of Fig.~\ref{fig:isolation_anti}), and yields an effective cavity coupling between the qubit and tunable resonator of $g/2\pi = 9.8$ MHz.
However, note that according to Eqs.~(\ref{eq:beta})--(\ref{eq:Meff}), $M_{eff}$ actually crosses through zero at $\Phi_x^{QFP}=\Phi_0/2$, ensuring the existence of a maximally off bias point. 
When the QFP $x$ loop is biased to $\Phi_0/2$, no anticrossing is observed, which implies a limit on the interaction strength of $g/2\pi \le 18.8$ kHz from the frequency resolution of the measurement (bottom plot of Fig.~\ref{fig:isolation_anti}).

\begin{figure}[t]
	\begin{center}
		\includegraphics[width=\columnwidth]{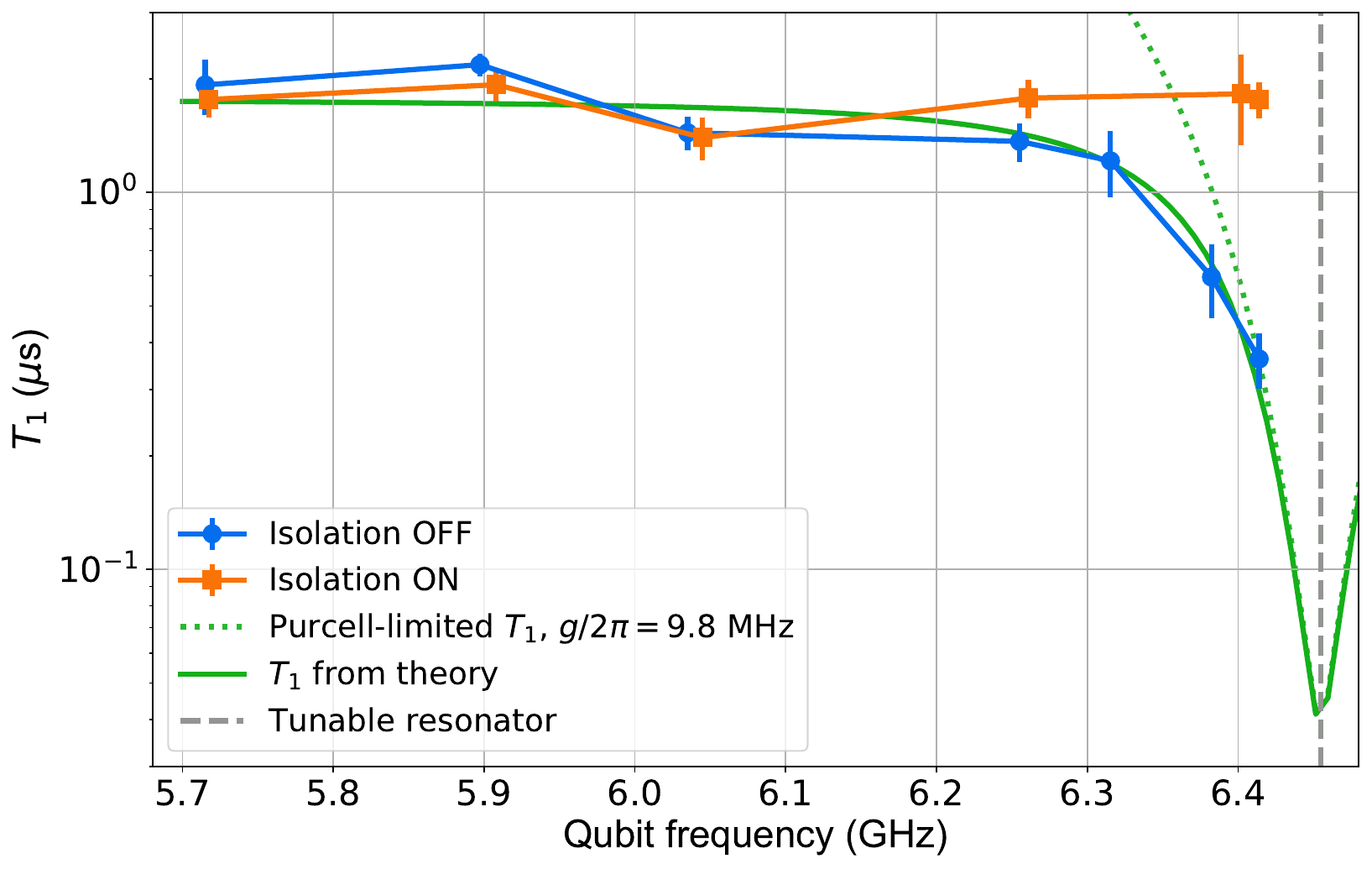}
	\end{center}
	\caption{
	Qubit lifetime as a function of frequency when the QFP isolates (orange squares) and does not isolate (blue circles) the qubit from the tunable resonator (frequency indicated by the gray dashed vertical line). 
	Lifetime values are determined by averaging over between 80 to 100 individual measurements, and the pictured error bars denote $1\sigma$.
	The calculated Purcell limited $T_1$ (green dotted trace) is reciprocally summed with the average isolated $T_1$ lifetime (green solid trace), and captures the behavior of the nonisolated flux qubit lifetime as the qubit is tuned toward the resonator.
	}
	\label{fig:isolation_t1}
\end{figure}

A similar effect can be seen in measurements of the flux qubit $T_1$ as a function of the QFP isolation (Fig.~\ref{fig:isolation_t1}).
The flux qubit is weakly coupled to a resonator at 7.19 GHz for the purposes of dispersive measurement of the excited state population.
In order to tune the frequency of the qubit between 5.715 and 6.414 GHz, $\Phi_x^{qub}$ is varied from $61.5$ to $63$ m$\Phi_0$ while $\Phi_z^{qub}$ is fixed at 0.
The tunable resonator frequency is marked by a dashed line at 6.46 GHz.
The average value of $T_1$ when the qubit is detuned far from the tunable resonator is $1.77$ $\mu$s, within the variance of previously published metrics for devices with similar characteristics ($I_p$ and shunt capacitance)~\cite{Yan2016}.
It is clear that, when the flux qubit is detuned by an amount greater than or equal to $400$ MHz from the tunable resonator (Fig.~\ref{fig:isolation_t1} grey dashed trace), the measured $T_1$ is insensitive to whether QFP isolation is on (Fig.~\ref{fig:isolation_t1} orange trace) or off (Fig.~\ref{fig:isolation_t1} blue trace).
In other words, in frequency ranges far detuned from $6.455$ GHz, qubit relaxation is dominated by other processes.
However, as the flux qubit is tuned toward the resonator frequency, the resonator becomes the dominant relaxation channel. 
When the QFP is maintained in the isolating state, the flux qubit $T_1$ is found to maintain its average value of about $1.77$ $\mu$s even within a few tens of megahertz of the resonator.
When isolation is turned off, the measured flux qubit $T_1$ demonstrates a rapid decline as the qubit approaches within $200$ MHz of the tunable resonator.
This behavior is demonstrated by the green solid trace in Fig.~\ref{fig:isolation_t1} and described by the equation $T_1 = (1/T_1^{avg}+1/T_1^{Purcell})^{-1}$, where $T_1^{avg}$ is the $1.77$ $\mu$s lifetime due to all other decay processes and $T_1^{Purcell}$ is the Purcell-limited lifetime calculated from Eq.~(4.65) of Ref.~\cite{Quintana2017phd}.
It is clear that the QFP protects the flux qubit from Purcell loss through the tunable resonator.

\section{CONCLUSIONS}
\label{sec:conclusions}
We experimentally verify a superconducting qubit readout technology that maintains speed and isolation from noise for a high-coherence flux qubit, requirements that are typically in tension.
We demonstrate readout of a flux qubit in the persistent current basis with separation fidelities surpassing $98.6\%$ with only $80$ ns of integration.
Integrating for $1$ $\mu$s increases the fidelity to $99.6\%$.
The use of the QFP as an amplifier allows for the flux-qubit-dependent frequency shift of the readout resonator to be more than nine linewidths, while only utilizing $2\%$ of the total qubit $z-$loop inductance for the readout mutual inductance.
Given the already-large frequency shift, $Q_e$ could be further reduced while maintaining a state-dependent shift of many linewidths.
This reduction in $Q_e$ would increase readout speed without affecting readout contrast.

Though a quantum-limited amplifier [a Josephson traveling-wave parametric (JTWPA) amplifier~\cite{Macklin2015}; see Appendix~\ref{sec:fridge}] is present in the current readout chain, we estimate that it is not necessary to achieve these readout speed and fidelity results.
The QFP-supplied amplification produces such a large frequency shift in the tunable resonator that the inherent signal contrast is excellent.
Moreover, without the JTWPA we could apply a stronger drive tone to the tunable resonator to make up for the loss in amplification and achieve a similar signal-to-noise ratio in the same integration time.
Eliminating quantum-limited amplifiers will reduce cost and complexity as high-coherence annealers continue to scale.

The limiter on readout fidelity is currently the ratio of the state-dependent flux shift in the QFP, $\Delta\Phi^{qub}$, to the QFP transition width, $w$.
Therefore, readout fidelity would be enhanced by increasing the mutual inductance between the flux qubit and QFP and/or decreasing the noise on $\Phi_z^{QFP}$ such that the transition becomes narrower.
For example, using $4\%$ of the qubit inductance for the transformer to increase the mutual inductance by a factor of 2 results in a $1$ m$\Phi_0$ wide operating region in which the readout fidelity is greater than $99.999\%$.

The QFP also provides isolation from the low-$Q$ resonator that enables the fast readout speed.
We have shown that the qubit lifetime does not degrade even as the qubit frequency is tuned to within $40$ MHz of the lossy resonator, confirming that the QFP protects the qubit from Purcell decay.
It is important to note that, since the frequency of the resonator is shifted by dc flux from the QFP, the qubit and resonator need not be proximate in frequency space to perform high-fidelity readout.

In light of the above, we stress that the demonstrated qubit readout scheme has opened the design space for both flux qubits and resonators.
Typical fabrication variations can lead to frequency crowding of resonators and to qubit-resonator collisions, problems that become especially pernicious in large systems with dispersive readout and/or minimally tunable qubits~\cite{Brink2018}.
The selective coupling/decoupling provided by the QFP permits one to more flexibly arrange qubits and resonators in frequency space to enable, e.g., multiplexed readout~\cite{Heinsoo2018}.
Normally qubit-resonator frequencies could collide and render the qubit useless unless tuned away from its flux-insensitive sweet spot, but this readout technology allows these frequencies to nearly overlap without detriment.
Additionally, direct resonator frequency collisions can be avoided by appropriate modulation of tunable resonator controls~\cite{Whittaker2016}. 
It is also feasible to develop calibration protocols for device characterization using only the tunable resonator, obviating the need to include any dispersively coupled resonators.
This provides a clear path to faster readout by suppressing unwanted qubit-resonator interactions and increasing feedline coupling to the resonator while simultaneously liberating the overall architecture from some of the onerous design constraints typical in conventional superconducting qubit systems.

Extensions beyond the standard operation outlined in this work could realize QND measurements~\cite{Schondorf2020, Wang2019} or measurements in the middle of an annealing sequence, which could be used to perform quantum simulation experiments on large systems of qubits~\cite{Harris2018, King2018, King2019}.

This work shows that fast, high-fidelity readout of low-$I_p$ flux qubits is possible without sacrificing qubit lifetime, out to the microsecond level.
We have identified multiple viable paths for its further optimization.
Our methodology is compatible with previously implemented scalable readout schemes~\cite{Whittaker2016}, and can even be extrapolated to improve conventional cQED readout of other qubit types such as transmons~\cite{Koch2007} and fluxonium~\cite{Manucharyan2009, Nguyen2019} by performing high-fidelity mapping from persistent-current states to energy eigenstates~\cite{Quintana2017phd}.
The combination of fidelity, speed, and isolation make it a critical enabling component of high-coherence quantum annealers.

\begin{acknowledgments}
We thank the Quantum Enhanced Optimization team for collaboration, especially the team at MIT Lincoln Laboratory for very helpful discussions, collaboration, and support, including Jonilyn Yoder, Steven Weber, Jamie Kerman, David Kim, George Fitch, Bethany Niedzielski Huffman, Alexander Melville, Jovi Miloshi, Arjan Sevi, Danna Rosenberg, Gabriel Samach, Cyrus Hirjibehedin, Simon Gustavsson, and William Oliver.
The research is based upon work supported by the Office of the Director of National Intelligence (ODNI), Intelligence Advanced Research Projects Activity (IARPA) and the Defense Advanced Research Projects Agency (DARPA), via the U.S. Army Research Office Contract No. W911NF-17-C-0050.
The views and conclusions contained herein are those of the authors and should not be interpreted as necessarily representing the official policies or endorsements, either expressed or implied, of the ODNI, IARPA, DARPA, or the U.S. Government.
The U.S. Government is authorized to reproduce and distribute reprints for Governmental purposes notwithstanding any copyright annotation thereon.
\end{acknowledgments}
\appendix
\section{EXPERIMENTAL SETUP AND WIRING}
\label{sec:fridge}
\begin{figure*}[t]
	\begin{center}
		\includegraphics[width=0.9\textwidth]{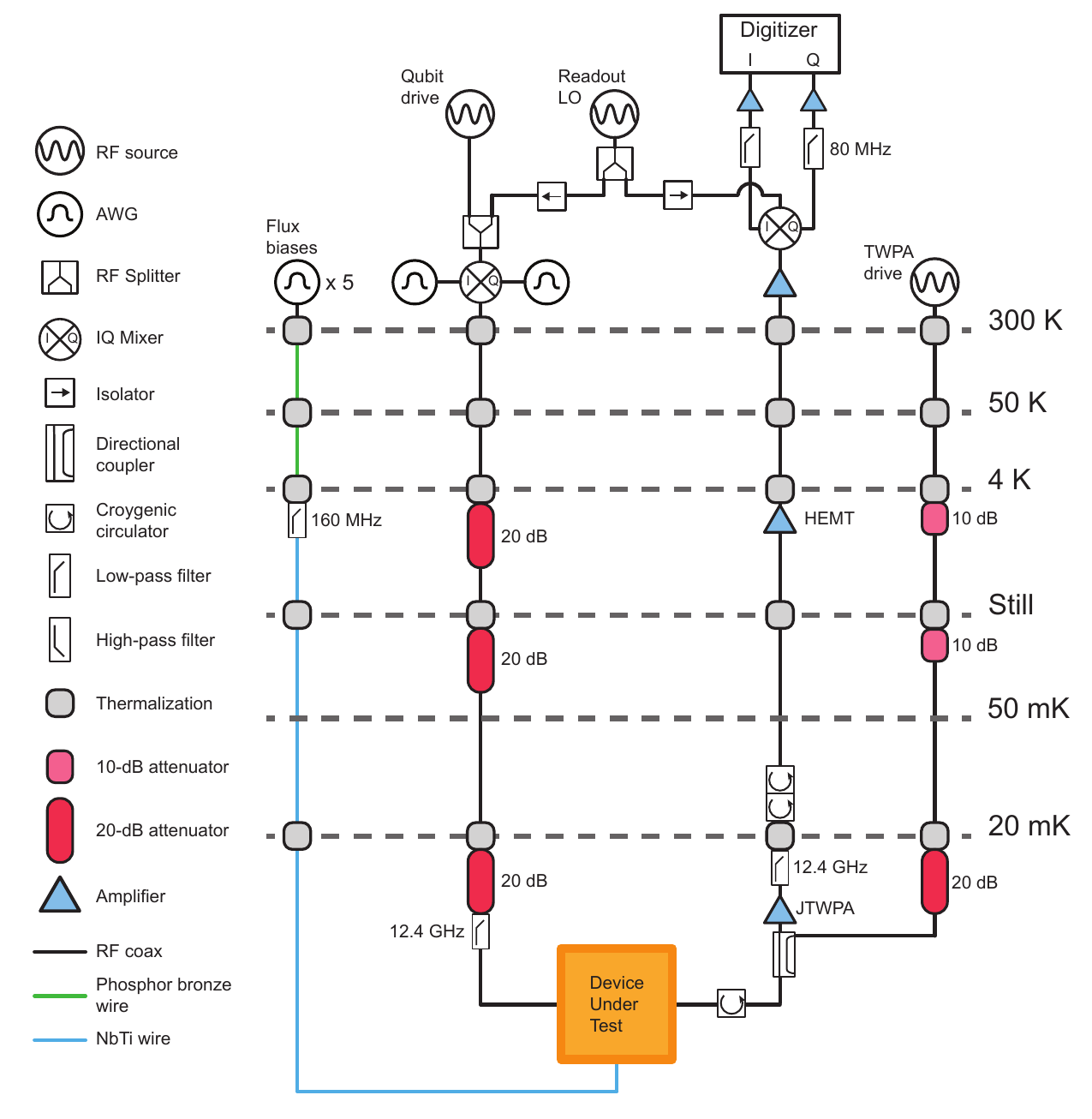}
	\end{center}
	\caption{
	A schematic diagram of the room-temperature measurement setup and dilution refrigerator wiring.
	}
	\label{fig:fridge}
\end{figure*}

In Fig.~\ref{fig:fridge} we present a schematic representation of the measurement setup from room temperature down to the mixing chamber.
The experiments are performed in a Leiden Cryogenics dilution refrigerator, with a base temperature in the range of $15-25$ mK.
Individual flux biases are provided by independent arbitrary waveform generator (AWG) channels and reach the device via phosphor bronze ribbon cables from 300 to 4 K, followed by NbTi cables from 4 K to the mixing chamber.
The isolation measurements presented in the main text are carried out with a slightly modified flux-bias configuration.
In addition to the displayed dc biases, high-frequency control is provided by different, independent AWG channels, utilizing their full 1 GS/s time resolution.
These rf biases are sent down the coax lines, and they are combined with the dc biases at the mixing chamber via cryogenic bias tees with an added 1-GHz low-pass filter.

Output signals are first amplified by a (JTWPA)~\cite{Macklin2015} at the mixing chamber, followed by a high-electron-mobility transistor (HEMT) amplifier anchored at $4$ K.
The JTWPA is included as a standard element in the dilution refrigerator wiring in order to enable the widest possible range of experiments. 
However, if optimizing for this particular type of readout, the JTWPA is likely not necessary, as it is expected to limit the total power used to drive the tunable resonator.
At high drive power, the tunable resonator is expected to provide similar signal to noise at integration times even shorter than the minimum presented in the main text.

We use a split-heterodyne configuration with image rejection to down-convert signals into the IF band, typically 50 MHz.
A field programmable gate array (FPGA) digitizer performs analog-to-digital conversion for signal processing and analysis.
A simple boxcar windowing function is applied for IF demodulation~\cite{Krantz2019}, which is done either directly on the FPGA or in software after the full signal traces are transferred off the card.
We did not perform further optimization of the signal integration, such as using nonlinear filter functions~\cite{Gambetta2007}.
Similarly, state discrimination relied on simple thresholding in the IQ plane, rather than machine-learning-based methods that can lead to improved fidelity in some circumstances~\cite{Magesan2015, Martinez2020}.

\section{DEVICE}
\label{sec:device}

\begin{figure*}[t]
	\begin{center}
		\includegraphics[width=\textwidth]{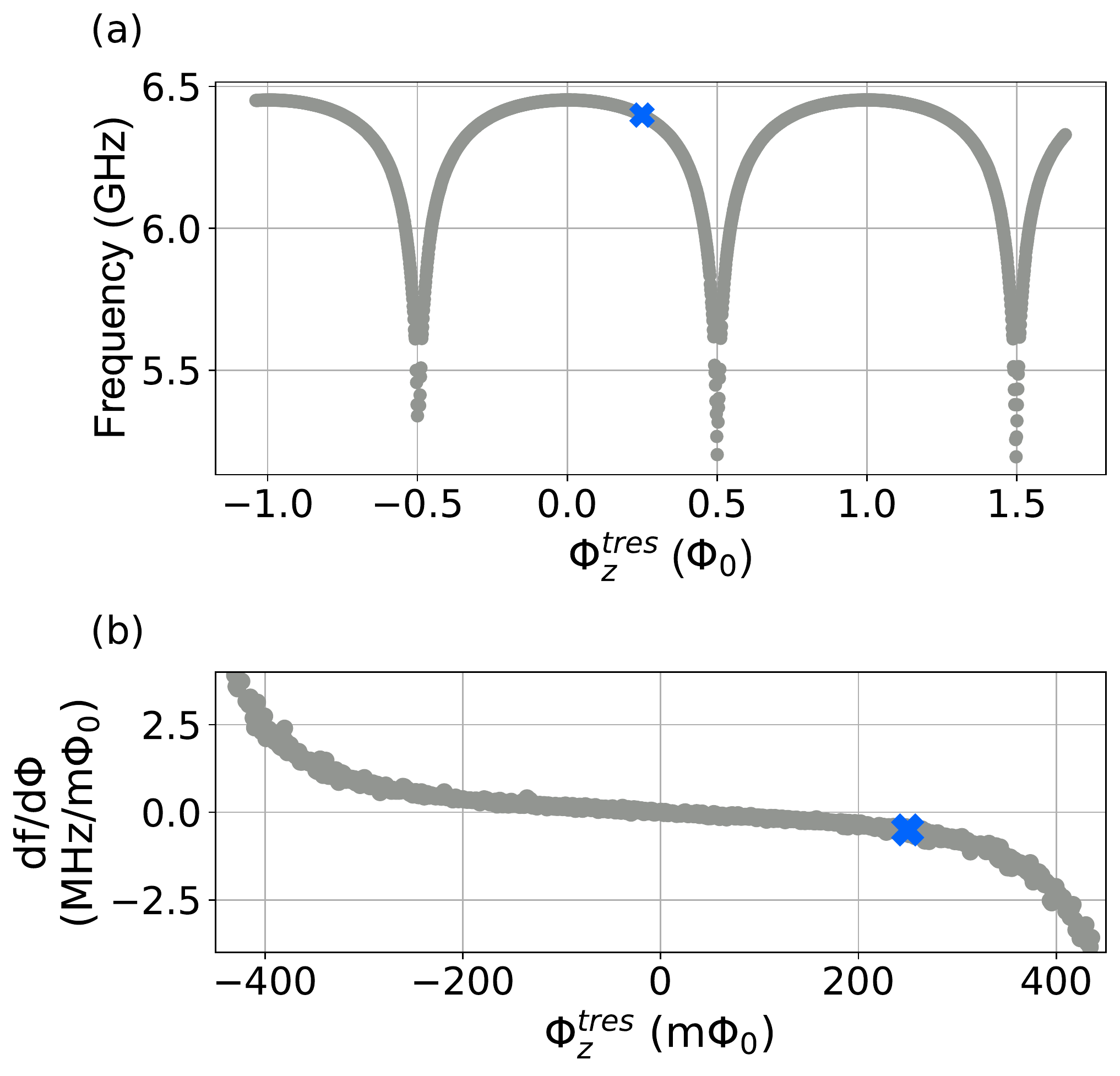}
	\end{center}
	\caption{
    An illustration of the physical device layout, generated by design software.
    The light pink regions delineate the high-quality MBE aluminum layer in which qubit capacitors, bias lines, resonators, and the ground plane are patterned.
    The blue regions depict the shadow-evaporated aluminum layer used to define Josephson junctions and SQUID loops.
    The purple bands represent aluminum air bridge crossovers used both within SQUID loops, as well as to connect different regions of the ground plane.
	}
	\label{fig:device}
\end{figure*}

In Fig.~\ref{fig:device} we show the layout of the device rendered by our design software.
The light pink regions delineate patterned high-quality molecular-beam epitaxy (MBE) aluminum used to define qubit shunt capacitors, bias lines, resonator waveguides, and the ground plane.
The blue regions depict aluminum patterned by shadow evaporation; these form the Josephson junctions and SQUID loops of the device.
Lastly, the purple bands represent aluminum air bridge crossovers~\cite{Chen2014, Rosenberg2020}.
These are used to create mutual inductive couplings between the qubit, the QFP, and the tunable resonator (concentric blue squares in the figure).
Furthermore, the crossovers help connect different regions of ground plane, suppressing spurious modes, and reducing bias line crosstalk.
The qubit, QFP, and tunable resonator are each delineated, and the large loops that form the mutual inductances are clearly visible.
Coplanar waveguides that enter from the edges of the image serve as flux control lines or the dispersive resonators used in calibration.
At the top left of the image, we note the two parallel rectangles that are the start of the large qubit shunt capacitor paddles.

In Table~\ref{tab:device} we present the target device parameters as designed, compared to some of the values as fabricated.
The models and methods for determining the qubit parameters are outlined in Ref.~\cite{Khezri2020}.
We find that some of the qubit parameter extraction methods are less amenable to determining the same parameters in the QFP.
However, the QFP readout margins are such that knowing them within fabrication tolerance is enough for achieving high performance.

\begin{table*}[ht]
    \begin{center}
    \begin{tabular}{lrr}
    \hline
    Parameter (units) & Designed & Extracted \\
    \hline
    \multicolumn{3}{l}{\textbf{Qubit}} \\
    x-loop junction critical current, $I_{c_{x}}^{qub}$ (nA) & 90 & 103 \\
    x-loop junction asymmetry, $d$ & 0 & 0.102 \\
    z-loop junction critical current, $I_{c_{z}}^{qub}$ (nA) & 194 & 228 \\
    Shunt capacitance, $C_{shunt}^{qub}$ (fF) & 47 & 70 \\
    Linear z-loop inductance, $L^{qub}$ (pH) & 133 & \\
    \hline
    \multicolumn{3}{l}{\textbf{QFP}} \\
    x-loop junction critical current, $I_{c_{x}}^{QFP}$ (nA) & 990 & \\
    x-loop junction asymmetry, $d$ & 0 & \\
    z-loop linear inductance, $L^{QFP}$ (pH) & 416 & \\
    Mutual inductance between qubit and QFP, $M^{qub,qfp}$ (pH) & 65 & \\
    Mutual inductance between QFP and tunable resonator, $M^{qfp,tres}$ (pH) & 65 & \\
    \hline
    \multicolumn{3}{l}{\textbf{Tunable Resonator}} \\
    z-loop critical current, $I_{c_{z}}^{tres}$ (nA) & 1200 & \\
    z-loop linear inductance, $L^{tres}$ (pH) & 199 & \\
    Total quality factor, $Q$ & 650 & 720\\
    \hline
    \end{tabular}
    \caption{
    Summary of designed device parameters, alongside some of the extracted parameters.
    }\label{tab:device}
    	\end{center}
\end{table*}

\section{TUNABLE RESONATOR CHARACTERIZATION}
\label{sec:tres}
\begin{figure}[t]
	\begin{center}
		\includegraphics[width=\columnwidth]{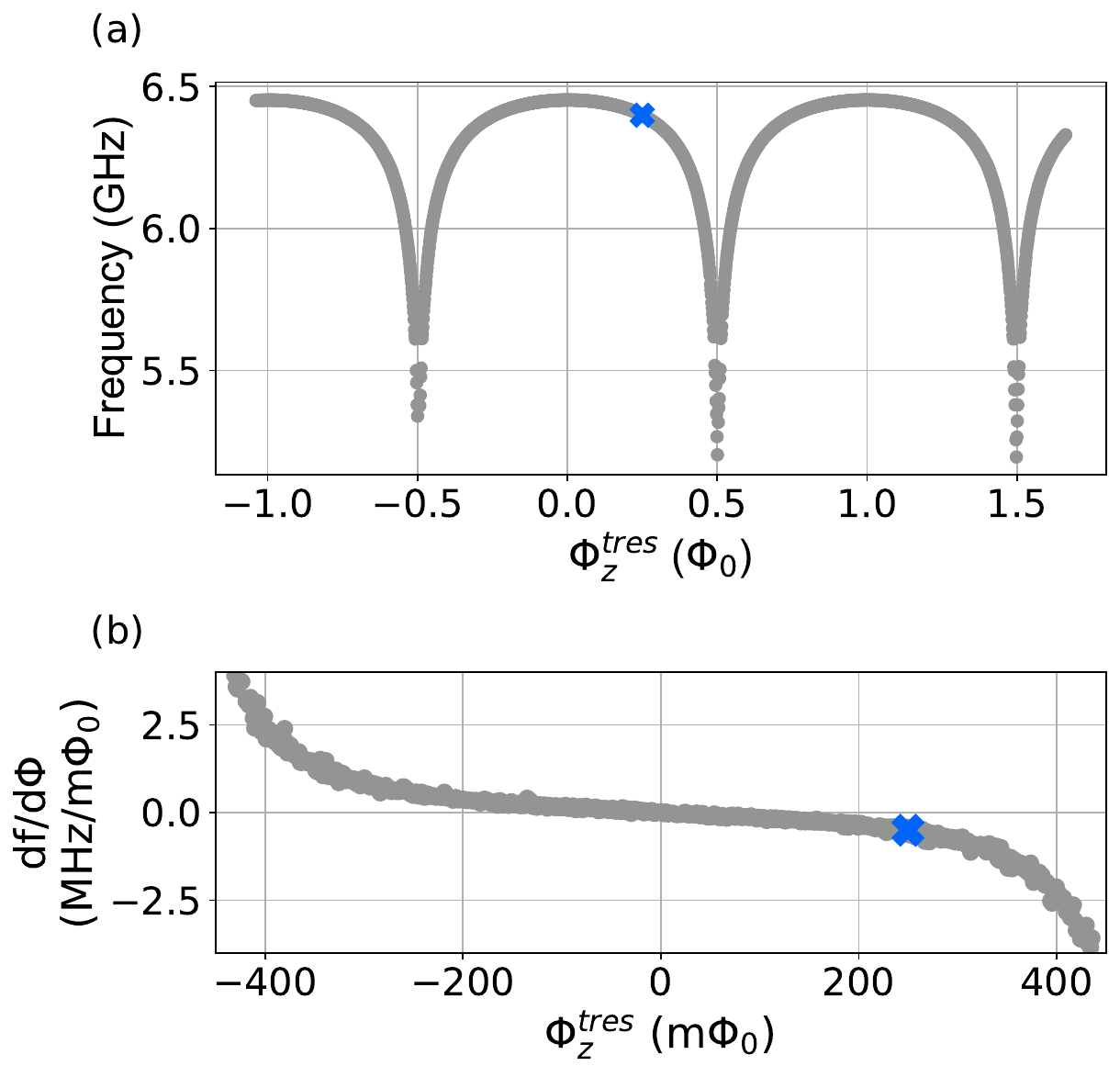}
	\end{center}
	\caption{
	(a) Tunable resonator frequency as a function of applied flux, where the frequencies are extracted from $S_{21}$ data.
	The full range of tunability is roughly $1.2$ GHz.
	The blue cross marks a typical operating point at $\Phi_z^{tres} = \Phi_0/4$.
	(b) The derivative of the tunable resonator modulation curve in (a) as a function of applied flux, in units of MHz/m$\Phi_0$.
	The flux range has been restricted to show a region within only $1$ $\Phi_0$ of flux tuning.
	The blue cross marks the same typical operating point as in the top plot.
	}
	\label{fig:modulation}
\end{figure}

As shown in Fig.~\ref{fig:schematic}, the tunable resonator is formed by a RF SQUID attached to the current antinode of a $\lambda/4$ resonator.rf
The flux applied to the rf SQUID modulates the frequency of the resonator, and the fabricated device achieves over 1 GHz of tunability (see Fig.~\ref{fig:modulation}(a)).
We operate the resonator close to an applied flux of $\Phi_{z}^{tres} = \Phi_0/4$, so that the resonator is more sensitive to changes in state of the QFP (see Fig.~\ref{fig:modulation}(b)).
Moving too close to $\Phi_0/2$ results in nonlinear resonator behavior at relatively low levels of microwave drive.
Further optimization of readout performance could be realized by more finely tuning the point that maximizes both the frequency shift of the resonator in addition to its power handling.

\begin{figure}[t]
	\begin{center}
		\includegraphics[width=\columnwidth]{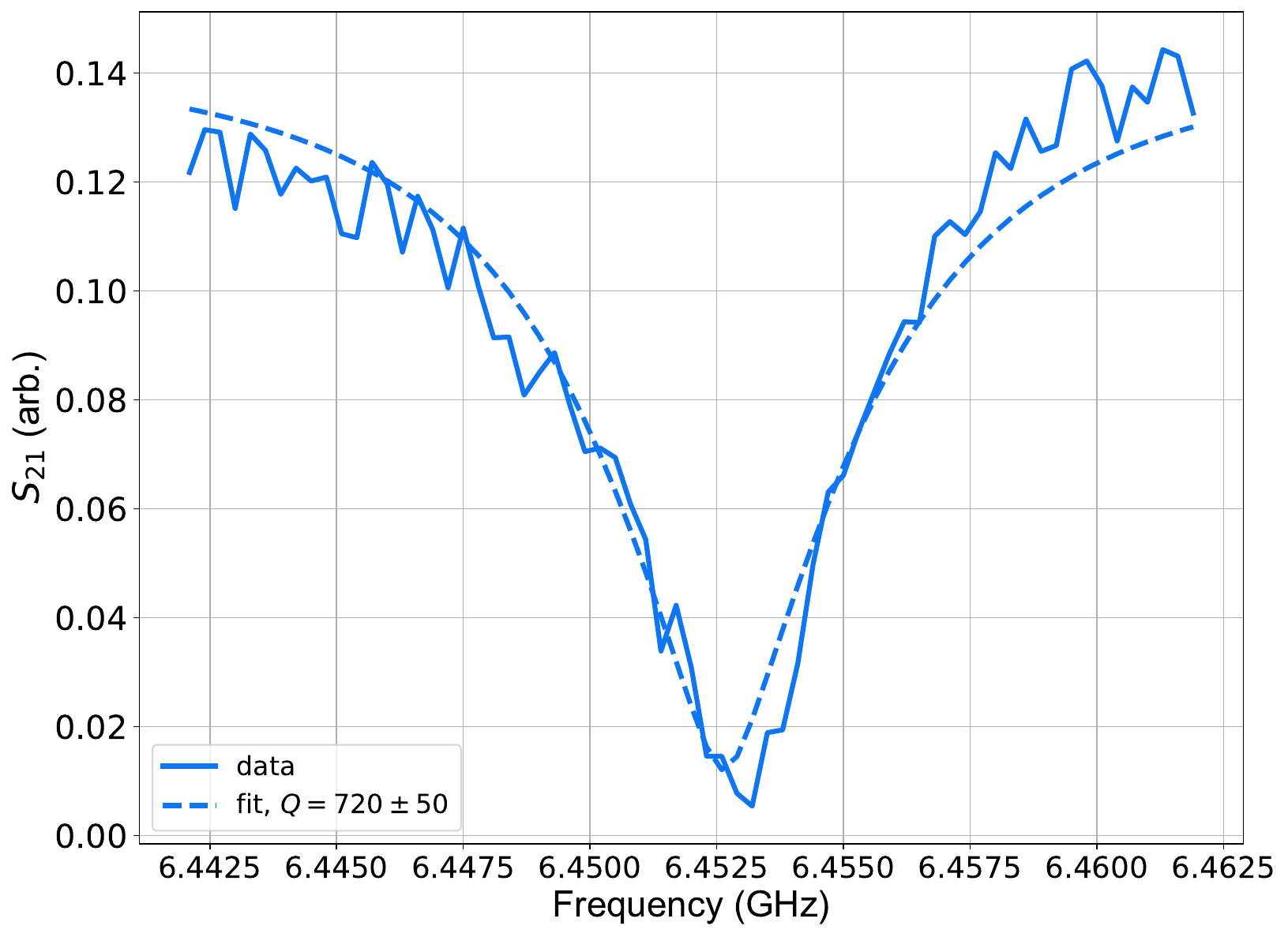}
	\end{center}
	\caption{
	Transmission and fit for the tunable resonator at its zero-flux bias point.
	The fit yields a total quality factor of $Q = 720 \pm 50$.
	}
	\label{fig:tresfit}
\end{figure}

To determine parameters of the resonator, we fit a simple model to the magnitude of transmission that accounts for asymmetry in the lineshape~\cite{Khalil2012}:
\begin{equation}
	| S_{21} | = A \left|1- \frac{\left(Q / \tilde Q_e \right)\, e^{i\phi}}{1 + 2 \, i \, Q \, \left(\frac{f-f_0}{f_0} \right)} \right|\,.
	\label{eq:s21}
\end{equation}
Here the external quality factor is related to the fit parameters by $Q_e = \tilde Q_e / \cos(\phi)$, and the internal quality factor is the usual $1/Q_i = 1/Q - 1/Q_e$.
The fit (see Fig.~\ref{fig:tresfit}) yields quality factors of $Q = 720 \pm 50$ and $Q_e = 760 \pm 120$, where the uncertainty is found via bootstrapping.

To further illustrate how large readout contrast is realized, see Fig.~\ref{fig:schematic}(c), which displays two transmission curves of the resonator, each for a single value of applied flux.
They correspond to a typical operating point of the resonator, where $\Phi_{z}^{tres} = \Phi_0/4$.
The light blue solid curve maps to the left circulating current state of the QFP.
When the direction of the state in the QFP flips, the corresponding change in flux imparts a frequency shift to the resonator of about 85 MHz (light blue dashed).
Thus, if we park the readout tone to be in the trough of the resonator when the QFP is in the $|L\rangle$ state, a small (large) integrated readout signal will map the QFP to its $|L\rangle$ ($|R\rangle$) state.


\section{QUANTUM CIRCUIT MODEL}
\label{sec:circuitizer}
\begin{figure*}[t]
	\begin{center}
		\includegraphics[width=\textwidth]{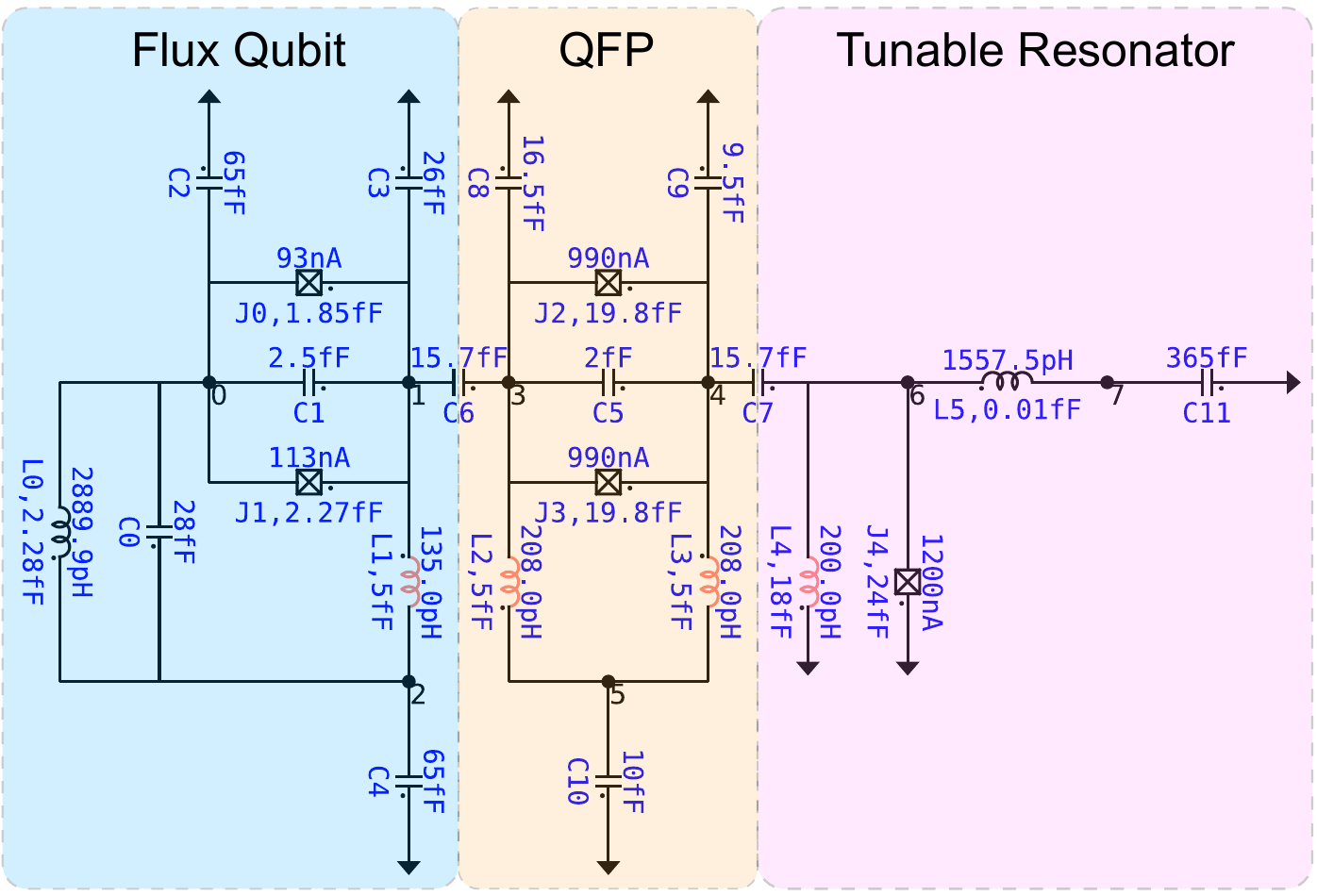}
    \end{center}
	\caption{
	Circuit schematic used to construct a Hamiltonian for quantum energy level calculations.
	The QFP is in the middle of the schematic and couples to the flux qubit and rf SQUID section of the tunable resonator by 60 pH mutual inductances at the red inductors.
	The parallel capacitances of the Josephson junctions and linear inductors are included with the junction and inductor symbols.
	The two junctions in the flux qubit $z$ loop are approximated by a linear inductor in order to reduce the dimensionality of the Hamiltonian (far left).
	The quarter-wave transmission line segment of the tunable resonator is approximated by a lumped-element model (far right).
	}
	\label{fig:circuitizer_schematic}
\end{figure*}

In Fig.~\ref{fig:circuitizer_schematic} we show the circuit schematic used to generate the dashed curves for comparison with the anticrossing data (see Fig.~\ref{fig:isolation_anti}).
The quantum energy levels are calculated using a combination of \textsf{QuTip}~\cite{Johansson2012, Johansson2013} and a Northrop Grumman proprietary Python package called \textsf{Circuitizer}.
\textsf{Circuitizer} is used to identify the normal modes of a circuit and generate the Hamiltonian in that basis.
In the case where $\Phi_x^{qub} = 0.626\,\Phi_0$, $\Phi_z^{qub} = -0.02\,\Phi_0$, and all other flux biases and charge offsets are set to 0, the explicit Hamiltonian is
\begin{widetext}
\begin{equation}
    \begin{split}
    \hat{H} = 1746.021 I& + 3.138 \hat{n}^{2}_{h2} + 5.331 \hat{n}^{2}_{h3} + 15.253 \hat{n}^{2}_{h4} + 27.121 \hat{n}^{2}_{h5} + 75.823 \hat{n}^{2}_{h6}\\
	& + 3.138 \hat{\theta}^{2}_{h2} + 5.331 \hat{\theta}^{2}_{h3} + 15.253 \hat{\theta}^{2}_{h4} + 27.121 \hat{\theta}^{2}_{h5} + 75.823 \hat{\theta}^{2}_{h6}\\
	& + (20.283+0.015i) e^{{-0.001}i\hat\theta_{h2}} e^{{-0.439}i\hat\theta_{h3}} e^{{0.023}i\hat\theta_{h4}} e^{{0.038}i\hat\theta_{h5}} e^{{0.160}i\hat\theta_{h6}}\\
	& - 491.717 e^{{-0.010}i\hat\theta_{h2}} e^{{-0.006}i\hat\theta_{h3}} e^{{0.270}i\hat\theta_{h4}} e^{{0.074}i\hat\theta_{h5}} e^{{-0.057}i\hat\theta_{h6}}\\
	& - 298.010 e^{{-0.030}i\hat\theta_{h2}} e^{{-0.001}i\hat\theta_{h3}} e^{{0.073}i\hat\theta_{h4}} e^{{-0.219}i\hat\theta_{h5}} e^{{0.034}i\hat\theta_{h6}}\\
	& + \text{H.c.}\,,
	\end{split}
    \label{eq:circutizer_hamiltonian}
\end{equation}
\end{widetext}
where $\hat{n}_{hi}$ and $\hat{\theta}_{hi}$ are the charge and phase operators on the $i^{th}$ normal mode, and $I$ is the identity.
In effect, the eight-node circuit is modeled with only five harmonic oscillator modes, since two modes are conserved charge modes and one is an oscillator mode of a frequency too high to affect the eigenenergies of the states of interest.
The five modes have dimensions of 14, 7, 3, 4, and 2, whose product makes the total Hilbert space dimension 2352. 
\textsf{QuTip} is used to solve for the eigenenergies of the Hamiltonian.
We determined that the precision of the lowest 14 energy states was better than $0.1$ MHz when that number of states is used to construct the Hamiltonian matrix.

\begin{figure*}[t]
	\begin{center}
		\includegraphics[width=\textwidth]{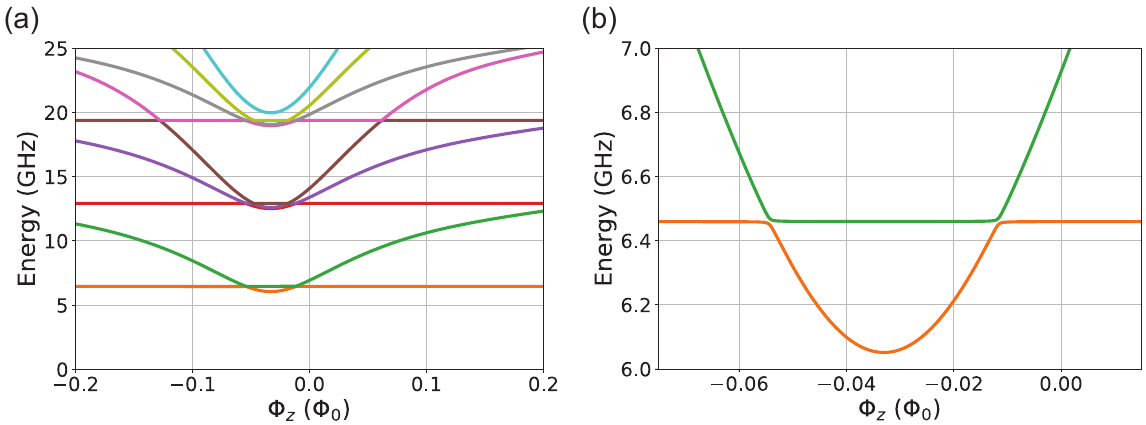}
    \end{center}
	\caption{
	Circuit energy levels calculated using QuTip on the Hamiltonian generated by \textsf{Circuitizer}.
	(a) Wide view of the lowest 10 energy levels as a function of $\Phi_z$.
	(b) Zoom-in of the first excited state of the flux qubit and its anticrossings with the fundamental mode of the tunable resonator.
	The energy levels are colored by energy ordering, not by mode, so the modes switch colors at the anticrossings.
	The shift of the minimum toward negative $\Phi_z^{qub}$ is caused by asymmetry in the $x-$loop junctions of the flux qubit.
	}
	\label{fig:circuitizer_z_sweep}
\end{figure*}

The circuit parameters are consistent with process control module (PCM) measurements, three-dimensional electromagnetic field models of the layout, and experiments presented in Ref.~\cite{Khezri2020}.
The critical currents of the flux qubit junctions are about 10\% higher than their design values, which is consistent with the wafer's critical current density as determined through room-temperature resistance measurements of PCM structures.
Each pair of adjacent red inductors has a mutual inductance of 60 pH between them.
High frequency structure simulator (HFSS) simulations of the layout predicted a mutual of 68 pH; however, the value in the \textsf{Circuitizer} model is within the margin of error given the extent of our knowledge of the material process.
The discrepancy could be due to the London penetration depth, $\lambda$, of the material being longer than that which was used in the HFSS model.
In Fig.~\ref{fig:circuitizer_z_sweep} we show the energy levels over a wider range of $\Phi_z^{qub}$ for a constant $\Phi_x^{qub} = 0.626\,\Phi_0$.
This is the bias configuration for the simulation that is compared with the anticrossing in the main text.
The shift of the energy levels toward negative $\Phi_z^{qub}$ is a result of an $x-$loop junction asymmetry of $10\%$ as defined in Ref.~\cite{Harris2009}.
In order to get the quantum model to overlay with good agreement on the anticrossing data in the main text, the $\Phi_z^{qub}$ and $\Phi_x^{qub}$ values need to be shifted by 4.4 and 6 m$\Phi_0$, respectively.
This is similar to the fit values determined in Ref.~\cite{Khezri2020} and is not unexpected, given the likelihood of offsets due to trapped flux in the chip.

\end{document}